\documentclass[aps,prl,twocolumn, superscriptaddress]{revtex4-1}

\usepackage{amsmath}
\usepackage{verbatim}
\usepackage{epsfig}

\newcommand{\om}{\omega}

\newcommand{\be}{\begin{equation}}
\newcommand{\ee}{\end{equation}}
\newcommand{\bea}{\begin{eqnarray}}
\newcommand{\eea}{\end{eqnarray}}

\begin{document}

\newcommand{\eg}{{\emph{e.g.} }}
\newcommand{\ie}{{\emph{i.e.} }}
\newcommand{\etal}{et al.}


\title{
Equivalence Principle and Gravitational Redshift}


\author{Michael A. Hohensee}
\email[]{hohensee@berkeley.edu}
\affiliation{Department of Physics, University of California, Berkeley, California 94720, USA}
\author{Steven Chu}
\altaffiliation{Present address: U.S. Department of Energy, 1000 Independence Avenue SW, Washington, D.C. 20585, USA}
\affiliation{Department of Physics, University of California, Berkeley, California 94720, USA}
\author{Achim Peters}
\affiliation{Institut f\"ur Physik, Humboldt-Universit\"at zu Berlin, Newtonstr. 15, 12489 Berlin, Germany}
\author{Holger M\"uller}
\affiliation{Department of Physics, University of California, Berkeley, CA 94720, USA}


\date{\today}

\begin{abstract}
We investigate leading order deviations from general relativity that violate the Einstein equivalence principle in the gravitational standard model extension. We show that redshift experiments based on matter waves and clock comparisons are equivalent to one another.  Consideration of torsion balance tests, along with matter wave, microwave, optical, and M\"ossbauer clock tests, yields comprehensive limits on spin-independent Einstein equivalence principle-violating standard model extension terms at the $10^{-6}$ level.
\end{abstract}

\pacs{}

\maketitle

Gravity makes time flow differently in different places.  This effect, known as the gravitational redshift, is the original test of the Einstein equivalence principle (EEP)~\cite{Einstein1911} that underlies all of general relativity; its experimental verification~\cite{PoundRebka,Vessot,redshift,Chou,Poli} is fundamental to our confidence in the theory.  Atom interferometer (AI) tests of the gravitational redshift~\cite{redshift,Poli} have a precision 10 000 times better than tests based on traditional clocks~\cite{Vessot}, but their status as redshift tests has been controversial~\cite{comment}.  Here, we show that the phase accumulated between two atomic wave packets in any interferometer equals the phase between any two clocks running at the atom's Compton frequency following the same paths, proving that atoms are clocks.  For a quantitative comparison between different redshift tests, we use the standard model extension (SME)~\cite{ColladayKostelecky,KosteleckyGravity,KosteleckyTassonPRL,KosteleckyTasson}, which provides the most general way to describe potential low energy Lorentz symmetry-violating (thus EEP-violating) signatures of new physics at high energy scales.  We show that all EEP tests are sensitive to the same five terms in the minimal gravitational SME~\cite{KosteleckyGravity,KosteleckyTassonPRL,KosteleckyTasson} and, for the first time, comprehensively rule out EEP violation in redshift tests greater than a few parts per million for neutral matter.

If two clocks are located at different points in spacetime, they can appear to tick at different frequencies, despite having the same proper frequency $\om_0$ in their local Lorentz frames. For clocks moving with nonrelativistic velocities $\vec{v}_{1}$ and $\vec{v}_{2}$ in a weak gravitational potential $\phi_{i}=-MG/|\vec{r}_{i}|$, the difference frequency is  \cite{MTW}
\be
\frac{\delta \om}{\om_{0}}\equiv \frac{\om_{1}-\om_{2}}{\om_0} =\frac{\phi_{1}-\phi_{2}}{c^{2}}-\frac{v_{1}^{2}-v_{2}^{2}}{2c^{2}}+O\left(c^{-3}\right).\label{eq:domoverom}
\ee
The first term is the gravitational redshift, originally measured~\cite{PoundRebka} by Pound and Rebka in 1960, while the second term is the time dilation due to the clocks' relative motion. The redshift term can be isolated from the time dilation if the clocks' trajectories are known.

The state of each clock can be described by a time-varying phase. If two clocks 1 and 2 are synchronized to have identical phase $\varphi_0=0$ at time $t=0$, then their relative phase $\delta \varphi_f\equiv\varphi_{1}-\varphi_{2}= \int \delta \om \,dt$ after a time $T$ is
\begin{equation}
\delta\varphi_f=\om_0\int_{0}^{T}dt\,\left(\frac{\vec r_{12} \cdot \vec g}{c^{2}}-\frac{v_{1}^{2}-v_{2}^{2}}{2c^{2}}\right),\label{eq:freeevrocket}
\end{equation}
specializing to a homogenous gravitational field so that $\phi_{1}-\phi_{2}=\vec g\cdot \vec r_{12}$, with $\vec r_{12}$ being the clocks' distance vector and $\vec{g}$ the local acceleration of free fall.  If the clocks are freely falling, then their motion is an extremum of their respective actions~\cite{MTW} $S_{i}=\int mc^{2}\, d\tau\approx\int  m[c^2+\phi_{i}-v_{i}^{2}/2]dt$.  Thus $\delta \varphi_f$ is proportional to the difference $S_{1}-S_{2}$ in their extremized actions.  
\begin{figure}
\centering
\epsfig{file=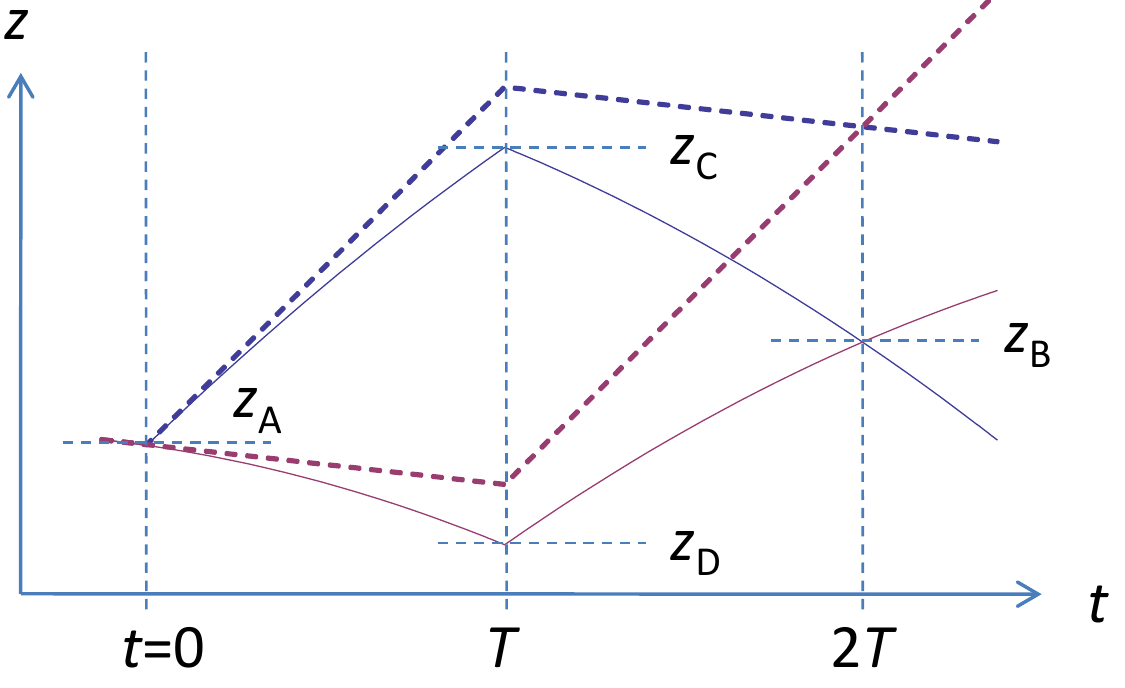,width=0.3\textwidth}
\caption{\label{MZ} Mach-Zehnder clock or atom interferometer.  Two otherwise freely falling clocks (or halves of an atomic wavepacket) receive momentum impulses that change their velocity by $\pm v_{r}$. The dashed lines indicate trajectories without gravity. }
\end{figure}

To clarify the equivalence between matter-wave and clock comparison tests, consider two conventional clocks that follow the two piecewise freely falling trajectories indicated in Fig.~\ref{MZ}.  
Initially, they are colocated and synchronized with zero phase difference.  
In a uniform gravitational field, it can be shown that the relative phase $\delta\varphi_{f}$ accumulated by the clocks in free fall vanishes, as the redshift and time dilation contributions in Eq.~(\ref{eq:freeevrocket}) are of the same magnitude $\omega_{0}gv_{r}T^{2}/c^{2}$ but opposite sign~\cite{Storey:1994}.  
Thus $\delta\phi_f=0$ is the measured phase difference at $t=2T$, when they are again colocated and at rest relative to one another.  
The problem can also be solved from the viewpoint of the moving clocks, rather than that of the stationary observer: $\delta\varphi_f$ vanishes because the time dilation term cancels the phase acquired during impulsive accelerations $a$ at $t=0$, $T$, and $2T$.  
This is calculated using Eq.~(2) with $\vec{a}$ replacing $\vec{g}$ (see Ref.~[12], chapter 13, part 6), and is given by
\begin{equation}
\delta\varphi_a=\lim_{\epsilon\rightarrow 0}\om_0\int_{t}^{t+\epsilon} dt\, \left(\frac{\vec{r}_{12}\cdot\vec{a}}{c^{2}}-\frac{v^2}{2c^2}\right) = \om_0\frac{\vec{v}_{r} \cdot\vec{r}_{12}}{c^{2}},\label{eq:accelphase}
\end{equation}
where $\vec{v}_r=\int_t^{t+\epsilon} \vec{a}\,dt$ is the velocity change.  The analogy with the atom interferometer is completed if each clock subtracts such a locally observable phase $\delta\varphi_a$ relative to a clock resting at $z=0$ (measurable, \emph{e.g.}, via radio signals~\cite{Vessot}) from its own phase after it accelerates, the total phase difference between the clocks is
\be
\delta\varphi=\delta\varphi_f+\om_0\frac{v_{r}}{c^{2}}(z_{\rm C}+z_{\rm D}-z_{\rm A}-z_{\rm B})=\om_0\frac{v_{r}gT^{2}}{c^{2}}.\label{eq:intphase}
\ee

Calculation of the phase measured by a Mach-Zehnder AI, such as employed by M\"uller, Peters, and Chu~\cite{redshift}, proceeds identically~\cite{Storey:1994}:  At $t=0$, a pulse from two counterpropagating lasers coherently divides an atomic matter wavepacket of mass $m$ and initial momentum $\vec{p}_{0}$ into a superposition of two different momentum states $\vec{p}_{0}$ and $\vec{p}_{0}+\hbar \vec{k}$.  The states separate with velocity of $v_r=\hbar k/m$ as shown in Fig.~\ref{MZ}. In free fall, the wavepackets follow paths $i$ which extremize their respective actions $S_i$ while accumulating phase in their local rest frames at the Compton frequency $\omega_{C}=mc^{2}/\hbar$. The total free evolution phase  is $\varphi_{i}=S_{i}/\hbar$, and just as before, $\delta\varphi_f=0$.  The atoms also acquire a phase shift of $\delta\varphi_{i}^{\rm L}=\pm kz$ each time they interact with the lasers~\cite{Storey:1994}.  This motion-dependent phase shift is identical to Eq.~(\ref{eq:accelphase}) after substituting $v_{r}=\hbar k/m$ and $\om_{0}=\omega_{C}$. 
Thus the phase difference in an atom interferometer is equal to that measured by two observers holding clocks oscillating at the Compton frequency that are moved along the atoms' path if they subtract the phase shifts they accumulate relative to a fixed clock during their periods of acceleration. This equivalence holds for AIs of any geometry, since an arbitrary path can be approximated to any desired accuracy by piecewise freely falling paths punctuated by momentum transfers.  The case of atoms held stationary in traps may be treated by Eq. (\ref{eq:accelphase}) without taking $\epsilon \rightarrow 0$. 
Thus atoms are effectively clocks that oscillate at $\om_{C}$. 
\begin{table*}[t]
\centering
\caption{\label{resultstable} Sensitivity of redshift experiments. The EEP-violation signal for each experiment is given as a linear combination of SME-parameters. The observable for the Pound-Rebka M\"ossbauer test, {\em e.g.,} is $-1.1 {\rm GeV}^{-1} \alpha (\bar a^n_{\rm eff})_0 -1.1{\rm GeV}^{-1} \alpha (\bar a^{e+p}_{\rm eff})_0 +(-0.34+[-0.66]) (\bar c^n)_{00} +(-0.34+[-0.006])(\bar c^p)_{00}+ 0.0002(\bar c^e)_{00}$, with $\bar a^{e+p}_{\rm eff}=\bar a^{p}_{\rm eff}+\bar a^e_{\rm eff}$. The last column shows the measured value and $1\sigma$ uncertainty. Signals dependent on models for $\xi$ are in square brackets. Curly brackets mark expected limits.}
\begin{tabular}{c
cccccccc}\hline\hline
Method & 
$\alpha (\bar a^n_{\rm eff})_0$ & $\alpha (\bar a^{e+p}_{\rm eff})_0$ & $(\bar c^n)_{00}$ & $(\bar c^p)_{00}$ & $(\bar c^e)_{00}$ & limit \\ & 
 GeV & GeV & & & & ppm \\ \hline
M\"ossbauer Effect~\cite{PoundRebka} & 
 -1.072 & -1.072 & 0.3358-[2/3] & -0.3353-[0.006] & 0.0001826 & $1000\pm7600$\\
H-maser on rocket~\cite{Vessot} & 
 -1.072 & -1.072 & 0.3358 & 0.3353-[0.67] & 0.0001826-[1.3] & $2.5\pm70$ \\
Cs Fountain (\emph{proj.})~\cite{ACES} & 
 -1.072 & -1.072 & 0.3358+[0.40] & 0.34+[0.28] & 0.0001826-[1.3] & $\left\{2 \right\}$\\
Bloch oscillations~\cite{redshift,Clade} &
 0.1632 & -0.1580 & -0.05112-[0.0005] & 0.04940+[0.0010] & $0.00002690$ & $3\pm 1$ \\
Bloch oscillations~\cite{Poli} & 
 0.1492 & -0.1439 & -0.04673-[0.0006] & 0.04500+[0.0008] & $0.00002451$ & $0.16\pm 0.14$ \\
Cs interferometer~\cite{redshift} & 
0.1881 & -0.1835 & -0.05890-[0.0004] & 0.05739+[0.001] & $0.00003126$ & $0.007\pm0.007$ \\
Rb interferometer~\cite{Merlet}  & 
0.1632 & -0.1580 & -0.05112-[0.0005] & 0.04940+[0.001] & $0.00002690$ & $-0.004\pm 0.007$ \\
\hline\hline
\end{tabular}
\end{table*}

Although they operate the same way, AI and conventional clock tests of the gravitational redshift could be sensitive to different physics beyond the standard model. The Compton frequency depends primarily on the physics of neutrons and protons, as these form the bulk of the atom's rest mass, whereas modern atomic clocks are sensitive to the physics of bound electrons, as we see below. A rigorous comparison between these tests requires the use of a consistent, comprehensive, and predictive phenomenological framework applicable to all experiments. The minimal gravitational SME~\cite{ColladayKostelecky,KosteleckyGravity,KosteleckyTassonPRL,KosteleckyTasson} is just such a framework, and provides the most general way to describe potential low energy Lorentz- and EEP-violating signatures of new physics at high energy scales. It preserves desirable features such as energy-momentum conservation, observer Lorentz covariance, and renormalizability of the nongravitational interactions, and is in extensive use~\cite{datatables}. The SME is formulated from the standard model Lagrangian by adding all Lorentz- or $CPT$ violating terms that can be formed from known fields and Lorentz tensors.  Different EEP tests will couple to different combinations of gravitational SME parameters.  

Without loss of generality, we may choose coordinates such that light propagates in the usual way through curved spacetime. 
The effects of EEP violation are then described by the $\alpha(\bar a^w_{\rm eff})_\mu$ and $(\bar c^w)_{\mu\nu}$ coefficients, which vanish if EEP is valid. The superscript $w$ takes the values $e,n$ and $p$ indicating the electron, neutron, and proton, respectively.  The motion of a test particle of mass $m^{\rm T}$, up to $O(c^{-3})$, is that which extremizes the action~\cite{KosteleckyTasson}
\begin{multline}
S = \int m^{\rm T}c\left( \sqrt{-\left(g_{\mu\nu}+2\bar c^{\rm T}_{\mu\nu}\right)dx^{\mu}dx^{\nu}}\right.\\
\left.+\frac{1}{m^{\rm T}}\left(a^{\rm T}_{\rm eff}\right)_{\mu}dx^{\mu}\right),\label{eq:smeaction}
\end{multline}
where $(a^{\rm T}_{\rm eff})_0= (1-2\phi \alpha)(\bar{a}^{\rm T}_{\rm eff})_0$ and $(a^{\rm T}_{\rm eff})_j=(\bar a^{\rm T}_{\rm eff})_{j}$ in a static potential.  For composite particles with $N^{e}$ electrons, $N^{p}$ protons, and $N^{n}$ neutrons 
\be\label{effectiveac}
(\bar c^{\rm T})_{\mu\nu}= \frac{1}{m^{\rm T}}\sum_wN^w m^w (\bar c^w)_{\mu\nu},\:\: (a_{\rm eff}^{\rm T})_\mu =\sum_w N^w (a_{\rm eff}^w)_\mu.
\ee
The metric $g_{\mu\nu}$ may also be modified by particle-independent gravity-sector corrections, as well as the $(\bar c^{\rm S})_{\mu\nu}$ and $(\bar a^{\rm S})_{\mu}$ terms 
in the action of the gravitational source body.  For experiments performed in the Earth's gravitational field, we may neglect such modifications as being common to all experiments.  Here, we focus on an isotropic subset of the theory~\cite{KosteleckyTasson} and thereby upon the most poorly constrained flat-space observable $(\bar c^{w})_{00}$ terms and the $(\bar a_{\rm eff}^{w})_{0}$ terms, that are detectable only by gravitational experiments~\cite{ColladayKostelecky,KosteleckyTassonPRL}.  The other $\bar c^{w}-$ and $\bar a^{w}-$ terms are respectively best constrained by nongravitational experiments or enter the signal as sidereal variations suppressed by $1/c$ and are neglected here.  

Expanding Eq. (\ref{eq:smeaction}) up to $O(c^{-2})$, dropping constant terms, and redefining $m^{\rm T}\rightarrow m^{\rm T}[1+\tfrac 53 (\bar c^{\rm T})_{00}]$ yields
\begin{equation}\label{lagrangesimplified}
S=\int m^{\rm T} c^{2}\left(\frac{\phi}{c^{2}}\left[1-\tfrac{2}{3}\left(c^{\rm T}\right)_{00}+\tfrac{2\alpha}{m^{\rm T}}\left(\bar{a}^{\rm T}_{\rm eff}\right)_{0}\right]
-\frac{v^{2}}{2c^{2}}\right)dt,
\end{equation}
where $v$ is the relative velocity of the Earth and the test particle.  Thus, at leading order, a combination of $\left(\bar c^{\rm T}\right)_{00}$ and $\alpha\left(\bar{a}^{\rm T}_{\rm eff}\right)_{0}$ coefficients rescales the particle's gravitational mass relative to its inertial mass.

The $(\bar c^{w})_{00}$ also cause a position-dependent shift in the binding energy of a composite particle.  This shift arises at $O(c^{-4})$ in the expansion of Eq. (\ref{eq:smeaction}), taking the form $v^{2}\phi\left(\bar c^{w}\right)_{00}$~\cite{KosteleckyTasson}.  Although negligible below $O(c^{-4})$ for the motion of a single elementary particle, composite systems bound by nuclear or electromagnetic forces can develop large internal velocities that are largely independent of $\phi$.  The $v^{2}\phi\left(\bar c^{w}\right)_{00}$ terms represent variations of the bound particles' inertial mass, and thus of the binding energy of the composite particle with $\phi$, with $\frac{1}{m^{w}}\rightarrow\frac{1}{m^{w}}\left[1+3\phi+\frac{5}{3}(\bar c^{w})_{00}-\frac{13}{3}\phi(\bar c^{w})_{00}\right]$~\cite{KosteleckyTasson}.  For clocks referenced to a transition between two bound states, this manifests as an anomalous rescaling~\cite{KosteleckyTasson} of the redshift by a clock-dependent factor of $1+\xi_{\rm clock}$.  Energy conservation requires variations in a particle's mass defect to be balanced by a rescaling of its gravitational mass~\cite{Schiff}, producing an additional correction to its motion.  To leading order, the gravitational rescaling factor for an atom's electronic binding energy is $\xi_{\rm elec.}^{\rm bind}= -(2/3)\left(\bar c^{e}\right)_{00}$~\cite{KosteleckyTasson}.  Scaled by the ratio of the electronic mass defect to the total mass, this contributes a fractional shift in the overall mass of an atom with $Z$ protons and $A$ nucleons with average nucleon mass $\bar{m}$ of order $ Z^{2}R_{\rm E}\xi_{\rm elec.}^{\rm bind}/(\bar{m}A)\sim 10^{-7}\xi_{\rm elec.}^{\rm bind}$, which we neglect.  Contributions from the nuclear binding energy are much larger, since the nuclear binding energy represents between $0.1$ and $1\%$ of an atom's mass (see below).

We begin with an analysis of gravity-probe A.  This experiment compared a hydrogen maser on the ground to an identical one carried on a rocket along a ballistic trajectory \cite{Vessot}.  A first influence of EEP violation in this experiment arises through a change in the motion of an object used to map the gravitational potential $\phi$ as a function of position. The gravitational acceleration $g^{\rm T}$ of a test mass $m^{\rm T}$ is found by minimizing the action Eq.~(\ref{lagrangesimplified}),
\be\label{gt}
g^{\rm T} = g\left(1+\beta^{\rm T}\right), \quad
\beta^{\rm T}=\frac{2\alpha}{m^{\rm T}}(\bar a_{\rm eff}^{\rm T})_0-\frac 23 (\bar c^{\rm T})_{00},
\ee
where $(\bar a_{\rm eff}^{\rm T})_0$ and $(\bar c^{\rm T})_{00}$ are obtained from Eq.~(\ref{effectiveac}).  The test mass moves as if it were in the potential $\phi'=(1+\beta^{\rm T})\phi$.
We need not consider anomalies in the motion of the rocket, as these are removed by continuous monitoring of the rocket's trajectory.  EEP-violation also causes a position-dependent shift of Hydrogen's $^{3}S_{1/2}$ to $^{1}S_{1/2}$ hyperfine transition. The hyperfine splitting scales with the electron mass $m^e$ and the proton mass $m^p$ as $(m^e m^p)^2/(m^e+m^p)^3$.  In analogy with a previous treatment of the Bohr energy levels in hydrogen~\cite{KosteleckyTasson}, the hyperfine transition varies linearly with $\phi$ as
\begin{equation}
\xi_{\rm H}^{\rm hfs}=-\frac{2}{3}\frac{m^{p}\left(2\bar{c}^{e}_{00}-\bar{c}^{p}_{00}\right)+m^{e}\left(2\bar{c}^{p}_{00}-\bar{c}^{e}_{00}\right)}{m^{p}+m^{e}}.
\label{eq:hyperfinexi}
\end{equation}
Note that here it is not necessary to rescale the electrostatic interaction, as was done in~\cite{KosteleckyTasson}, since it cancels in the evaluation of $\xi$.  Expressed in terms of the potential $\phi'$, the signal becomes
\begin{equation}\label{eq:gpa}
\frac{\delta f}{f_{0}}=\frac{\phi'{}_{s}-\phi'{}_{e}}{c^{2}}\left(1+\xi_{\rm H}^{\rm hfs}-\beta^{\rm SiO_{2}}\right)-\frac{v_{s}^{2}}{2c^{2}}.
\end{equation}
The precise combination of $\left(\bar c^{w}\right)_{00}$ and $\left(\bar a^{w}_{\rm eff}\right)_{0}$ bounded by GP-A is given in Table~\ref{resultstable}.  For simplicity, we assume that the potential $\phi'$ has been mapped by test masses made of silicon dioxide.  This assumption is made throughout. 
A similar analysis applies to the Atomic Clock Ensemble in Space (ACES) mission~\cite{ACES}, which aims to place a Cs clock alongside a H-maser aboard the International Space Station.  In a rough hydrogenic model, we estimate $\xi_{\rm Cs}^{\rm hfs}$ using Eq.~(\ref{eq:hyperfinexi}) by replacing $(\bar c^{p})_{00}$ with $(\bar c^{\rm ^{133}Cs})_{00}$ from Eq.~(\ref{effectiveac}), and the proton mass $m^{p}$ with that of Cesium.  Thus, a comparison of onboard Cs clocks to those on the ground measures $\xi^{\rm hfs}_{\rm Cs}-\beta^{\rm SiO_{2}}$ (Table~\ref{resultstable}).

Null tests comparing clocks 1 and 2 with clock coefficients $\xi_{1}$ and $\xi_{2}$ as they move together through a gravitational potential can yield bounds~\cite{KosteleckyTasson} on $\xi_1- \xi_2$.  

The Pound-Rebka experiment~\cite{PoundRebka} measured the gravitational redshift of a $14.4$\,keV transition in stationary $^{57}$Fe nuclei.  
With $Z=26$, $^{57}$Fe has an unpaired valence neutron that makes a transition between different orbital angular momentum states.  Assuming the transition energy scales with the reduced mass of the neutron, the Pound-Rebka experiment constrains (Tab.~\ref{resultstable})
\begin{equation}\label{eq:mossb}
\xi_{^{57}{\rm Fe}}^{\rm Mossb.}-\beta^{\rm grav}=-\frac{2}{3}\frac{m^{^{56}{\rm Fe}}c^{n}_{00}+m^{n}c^{^{56}{\rm Fe}}_{00}}{m^{^{57}{\rm Fe}}}-\beta^{\rm grav}.\end{equation}

Determination of the EEP-violating phase in an AI proceeds along the same lines as the analysis leading up to Eq.~(\ref{eq:intphase}), substituting the EEP-violating action of Eq.~(\ref{lagrangesimplified}).
To leading order, we obtain $\delta \varphi=(1+\beta^{\rm At})k gT^2$.  This reproduces the result obtained in~\cite{redshift}, with $\beta^{\rm At}$ given by Eq.~(\ref{gt}) specific to the atomic species.  
AIs are also sensitive to variations in the atoms' binding energy resulting from changes to the inertial mass of their constituent particles.  
Estimates of $\xi_{\rm nuc.}^{\rm bind}$ are strongly model dependent. We derive the values in Table~\ref{resultstable} by treating the nucleus as a Fermi gas confined in a square potential well of constant radius, holding fixed the last nucleon's binding energy.  Thus we find that the AI constrains $\beta^{\rm At}+\xi^{\rm bind}-(\beta^{\rm grav}+\xi^{\rm grav})$, where $\xi^{\rm grav}$ is the small contribution of the gravimeter's binding energy to its motion (Tab.~\ref{resultstable}). Bloch oscillations \cite{Clade,Poli} are a special case of an AI with the atoms at rest and bound the same terms if they use the same species, see Tab.~\ref{resultstable}.

We have demonstrated that the phase difference measured in any AI is exactly the same as the phase accumulated by a pair of conventional clocks following the same path, ticking at $\om_{C}$.  Experiments on different particles or transitions offer windows on different sets of SME parameters. The sensitivities of various redshift tests to the $a-$ and $c-$ coefficients of the SME are summarized in Tab.~\ref{resultstable}.  In combination with the two best torsion balance tests of the universality of free fall (UFF)~\cite{Gundlach2009}, which limit $\beta^{\rm Be}+\xi^{\rm Be}-\beta^{\rm Ti}-\xi^{\rm Ti}$ and $\beta^{\rm Be}+\xi^{\rm Be}-\beta^{\rm Al}-\xi^{\rm Al}$, we obtain simultaneous bound on all five EEP violation parameters for normal, neutral matter, see Tab.~\ref{limits}.  This is the first time that each has been bounded without assuming the others are zero, closing any loopholes for renormalizable EEP violations for neutral particles at $O(c^{-2})$: Because the SME is comprehensive \cite{ColladayKostelecky,KosteleckyGravity,KosteleckyTasson}, any additional anomalies may require as yet unknown particles, violation of energy conservation, or other innovations. We have, however, assumed that particles mediating binding potentials (\emph{e.g.} W-bosons, $\pi$-mesons, etc.) satisfy EEP. AI and UFF tests have the best sensitivity to meson-related anomalies in the nuclear binding energy.  Spin-dependent anomalies are not observable by existing redshift tests, and are a promising area for future study.

\begin{table}
\centering
\caption{Limits ($\times10^6$), estimated by multivariate normal analysis \cite{Hogg} using 
tests listed in Tab. \ref{resultstable}, torsion balance tests~\cite{Gundlach2009}, and relative redshift measurements~\cite{KosteleckyTassonPRL,Blatt,Ashby,Fortier}, with $1\sigma$ uncertainties. The index $T$ replacing $0$ indicates these limits hold in the Sun-centered celestial equatorial frame \cite{datatables}.
\label{limits}}
\begin{tabular}{ccccc}\hline\hline
$\alpha (\bar a^n_{\rm eff})_T$ & $\alpha (\bar a^{e+p}_{\rm eff})_T$  & $(\bar c^n)_{TT}$ & $(\bar c^p)_{TT}$ & $(\bar c^e)_{TT}$ \\
(GeV) & (GeV)\\
$4.3\pm3.7$ & $0.8\pm1.0$ & $7.6\pm6.7$ & $-3.3\pm3.5$ & $4.6\pm4.6$\\
\hline\hline
\end{tabular}
\end{table}

Redshift and UFF tests differ in their style of execution, as the former compare proper times whereas the latter compare accelerations, but the EEP violations they constrain take the same form at $O(c^{-2})$, consistent with Schiff's conjecture.
EEP-violation entering at $O(c^{-4})$, however, may allow UFF to be valid at one point in a gravitational field, but be violated elsewhere; A single UFF test on the ground might not necessarily imply local position invariance (LPI)~\cite{TEGP}, and might thus need to be complemented by redshift measurements.  Future AIs may capable of constraining $O(c^{-4})$ physics~\cite{MikeCPT}.

We thank Q. G. Bailey, N. Gemelke, V. A. Kosteleck\'y, W. Schleich, 
 and J. Tasson for discussions and the David and Lucile Packard Foundation, the National Institute of Standards and Technology,
 and the Alfred P. Sloan Foundation for support.



\end{document}